\documentclass{article}

\usepackage{arxiv}

\usepackage[utf8]{inputenc} % allow utf-8 input
\usepackage[T1]{fontenc}    % use 8-bit T1 fonts
\usepackage{hyperref}       % hyperlinks
\usepackage{url}            % simple URL typesetting
\usepackage{booktabs}       % professional-quality tables
\usepackage{amsfonts}       % blackboard math symbols
\usepackage{nicefrac}       % compact symbols for 1/2, etc.
\usepackage{microtype}      % microtypography
\usepackage{lipsum}		% Can be removed after putting your text content
\usepackage{graphicx}
\usepackage{natbib}
\usepackage{doi}
\usepackage{float}

\newcounter{assumptioncounter}
\newcommand{\setassumptioncounter}{
    \refstepcounter{assumptioncounter}
    \renewcommand{\theequation}{A\arabic{assumptioncounter}}
}
\newcommand{\setequationcounter}{
    \renewcommand{\theequation}{\arabic{equation}}
}

\usepackage[figuresright]{rotating}
\usepackage{amsmath}
\usepackage{amssymb}
\usepackage{bbm}
\usepackage{titlesec}
\usepackage{caption}

\title{Estimation of Policy-Relevant Causal Effects in the Presence of Interference with an Application to the Philadelphia Beverage Tax}

\author{Gary Hettinger \\
	Division of Biostatistics\\
	University of Pennsylvania\\
	Philadelphia, PA, U.S.A. \\
	\texttt{ghetting@pennmedicine.upenn.edu} \\
	\And
	Christina Roberto \\
	Department of Medical Ethics and Health Policy\\
	University of Pennsylvania\\
	Philadelphia, PA, U.S.A. \\
        \And
	Youjin Lee$^*$ \\
	Department of Biostatistics\\
	Brown University\\
	Providence, RI, U.S.A. \\
        \And
	Nandita Mitra$^*$ \\
	Division of Biostatistics\\
	University of Pennsylvania\\
	Philadelphia, PA, U.S.A \\
}

\renewcommand{\shorttitle}{Estimation of Policy Effects with Interference}

\hypersetup{
pdftitle={Estimation of Policy-Relevant Causal Effects in the Presence of Interference with an Application to the Philadelphia Beverage Tax},
pdfauthor={Gary~Hettinger, Christina~Roberto, Youjin~Lee, Nandita~Mitra},
pdfkeywords={Difference-in-Differences, Doubly Robust, Health Policy, Spillover},
}

\begin{document}
\maketitle

\begin{abstract}
To comprehensively evaluate a public policy intervention, researchers must consider the effects of the policy not just on the implementing region, but also nearby, indirectly-affected regions. For example, an excise tax on sweetened beverages in Philadelphia was shown to not only be associated with a decrease in volume sales of taxed beverages in Philadelphia, but also an increase in sales in bordering counties not subject to the tax. The latter association may be explained by cross-border shopping behaviors of Philadelphia residents and indicate a causal effect of the tax on nearby regions, which may offset the total effect of the intervention. To estimate causal effects in this setting, we extend difference-in-differences methodology to account for such interference between regions and adjust for potential confounding present in quasi-experimental evaluations. Our doubly robust estimators for the average treatment effect on the treated and neighboring control relax standard assumptions on interference and model specification. We apply these methods to evaluate the change in volume sales of taxed beverages in 231 Philadelphia and bordering county stores due to the Philadelphia beverage tax. We also use our methods to explore the heterogeneity of effects across geographic features.
\end{abstract}

% keywords can be removed

\section{Introduction}
\label{sec:intro}

In January 2017, the City of Philadelphia, Pennsylvania (PA) implemented an excise tax of 1.5 cents per ounce on sugar- and artificially-sweetened beverages to raise revenue for educational initiatives including the city's Pre-Kindergarten expansion and Community Schools program (\cite{PhillyTaxCode}). The decision was motivated also in part by studies that associated excise taxes with reduced intake of taxed beverages (\cite{Brownell2009, CabreraEscobar2013}). City-level policy makers hoped to reduce the consumption of such beverages, given evidence linking sweetened beverage consumption to negative health outcomes such as obesity and type 2 diabetes (\cite{Hu2013}). 
%, both of which were on the rise in Philadelphia through 2017 (\cite{PhillyChart2019}). 
Despite generating over $\$330$ million dollars in revenue from January 2017 to June 2021 (\cite{RhynhartController2022}), there have been recent efforts to repeal the tax, motivated by claims of disproportionate economic burden and loss of retailer profits. On the other hand, several studies have shown the benefits of the Philadelphia beverage tax (PBT) in reducing sales, which presumably have led to a reduction in intake (\cite{Roberto2019, Lawman2020, Bleich2021, Edmondson2021, Petimar2022}). 

To assess the causal effects of public policies and excise taxes, researchers often use a difference-in-differences (DiD) approach, which estimates the effect of the intervention by taking the difference in outcome trends between comparable regions with and without the intervention of interest (\cite{Ashenfelter1978, Ashenfelter1984}). Previous studies employed DiD methods to compare Philadelphia (treated region) to Baltimore, Maryland (control region), finding that city-level sweetened beverage volume sales declined by $51\%$ in Philadelphia in the year following tax initiation (\cite{Roberto2019}) with evidence of sustained declines two years after tax initiation (\cite{Petimar2022}). 

Underlying the causal interpretation of the DiD framework are strong identification assumptions, including the key but untestable counterfactual parallel trends assumption, which necessitates that the average outcomes of the treated group would have evolved in parallel with the average outcomes of the control group had the intervention never occurred (\cite{Heckman1997}). Previous authors have developed methods to relax this assumption in order to estimate the average treatment effect on the treated (ATT), requiring that counterfactual parallel trends hold only after conditioning on observed pre-intervention confounding variables. \cite{Heckman1997} presented methods to adjust for confounding with outcome regression modeling or propensity score matching, requiring correctly specified outcome or propensity score models, respectively. \cite{Abadie2005} used a propensity score model to develop an inverse probability of treatment weighting (IPW) estimator. Recent work by \cite{Li2019} and \cite{Santanna2020} further relaxed model specification assumptions by developing doubly robust estimators for the ATT which require only that at least one of the outcome and propensity score models is correctly specified. 

Many policy evaluations face the added complexity that individuals may avert taxes and restrictions by crossing into neighboring regions where the policy is not in place. Evidence of this behavior is common in practice for excise and sales taxes (\cite{Asplund2007}), gun policies (\cite{Raifman2020}), marijuana restrictions (\cite{Hao2020}), and more. Evaluations of excise taxes on sweetened beverages have generally found evidence of significant cross-border purchasing, with a few exceptions (\cite{Andreyeva2022}). When studying the effects of the PBT, \cite{Roberto2019} and \cite{Petimar2022} used DiD methods to compare Philadelphia-bordering counties to Baltimore and estimated that $25-30\%$ of the total effect of the PBT on volume sales was offset by cross-border shopping. These behaviors fall under the umbrella labeled  \textit{interference} and violate the Stable Unit Treatment Value Assumption (SUTVA) from Rubin's formulation of the potential outcomes framework for causal inference (\cite{Rubin1980}). While the literature has synonymously referred to this violation as \textit{spillover}, we find it more intuitive to reserve that term for situations where a neighboring region experiences the direct extension of the effect on the intervened region, such as  when a vaccination mandate provides an additive layer of protection to nearby regions. Conversely, the desired policy effects in the aforementioned policy studies are likely reduced by individuals crossing regional boundaries to bypass impositions. Accordingly, here we call this subset of interference a \textit{bypass} effect. 

Major advances have been developed to identify and estimate direct and spillover effects in controlled and observational studies where interference is believed to occur between individuals of a particular group, but not across these groups (\cite{HudgensHalloran2008, Tchetgen2012, LiuDR2019, Papadogeorgou2019, Huber2021}). However, little has been published on DiD methods that specifically target bypass effects in policy evaluations, where an intervention is introduced to an entire group and bypass effects occur between groups. Working papers by \cite{Clarke2017} and \cite{Butts2021} have defined causal estimands of interest and identification assumptions for general potential outcomes under interference, but rely on a two-way fixed effects model (TWFE), which imposes strict parametric and effect homogeneity assumptions to identify an unbiased treatment effect (\cite{DeChaisemartin2022}). 

In this work, we develop flexible methodology to robustly estimate the causal effects of the PBT on both Philadelphia and its neighboring counties under interference while also adjusting for confounding. This methodology is doubly robust in that our estimators are consistent if at least one of the propensity score or outcome models, which can be estimated non-parametrically, are well-specified. Our ensuing analysis serves to provide a framework for practitioners studying policies susceptible to bypass effects as well as further evidence of the effect of the PBT on volume sales in Philadelphia and its neighboring regions.

In what follows, Section~\ref{sec:data} introduces the PBT study data. In Section~\ref{sec:methods}, we present relevant notation, review a potential outcomes framework under interference, propose a modified SUTVA, and present doubly robust estimators for the treated and bypass effects. In Section~\ref{sec:rda}, we conduct an analysis of the PBT to provide new insights on the comprehensive causal effect of the tax. We conduct simulation studies to compare the finite sample performance and empirically verify robustness properties of different DiD methods under realistic scenarios in Section~\ref{sec:sims}. We conclude with a discussion in Section~\ref{sec:disc}.

\section{Philadelphia Beverage Tax Data}
\label{sec:data}

Beverage price and sales data were purchased from the market research firm Information Resources Inc (IRI), which obtains data  from major US retailers described elsewhere (\cite{MuthIRI2016}). For our study, we used retail sales data reported in 4-week periods for beverages sold from January 1, 2016 to December 31, 2017 in stores from Philadelphia, other PA counties, and Baltimore. Baltimore was chosen to be a comparison city due to its demographic and geographic similarities to Philadelphia and was not directly or indirectly affected by a beverage excise tax. 
% Non-Border stores within 6 miles of the Philadelphia border were removed to ensure this group was not indirectly affected by the beverage tax. 
Data were provided at the individual beverage level based on a unique universal product code and aggregated at the store-level. Store and beverage categorization, as well as price and sales aggregations, were conducted as described in \cite{Roberto2019}. 

For each store, we observed  price and sales records at 26 time points (13 before and 13 after tax implementation) for multiple taxed and non-taxed beverage categories, which accounted for \$14.3 billion and \$19.9 billion in sales, respectively. We further merged store classifications into two categories: one encompassing supermarkets, grocery stores, and mass merchandisers (SGMs), and one for pharmacies which often demonstrate different consumer purchasing behaviors than SGMs. Among the 558 stores in our study, 180 were Philadelphia stores (40 SGMs, 140 pharmacies), 318 were stores from other PA counties (123 SGMs, 195 pharmacies), and 60 were Baltimore stores (15 SGMs, 45 pharmacies). We additionally linked zip code-level socioeconomic and racial census data from 2016 to each of the stores in our study. The IRI data  contained no missing data.

\section{DiD Methodology for Causal Effects on Treated and Neighboring Control Regions}
\label{sec:methods}

Here we develop doubly robust estimators for both the policy intervention effect and the bypass effect. To introduce our proposed framework, we first introduce a potential outcomes representation under interference, called an exposure mapping (\cite{Aronow2013}), and define relevant estimands under this representation. We then provide conditions necessary to identify these estimands and present our proposed DiD estimators.  

\subsection{Exposure Mapping for Potential Outcomes under Interference}
\label{sec:methods:exposure}

Assume we have a collection of units, $i=1,\dots,n$, observed across a pre- and post-treatment period, $t=0,1$. For simplicity, we will first introduce methodology in the setting with a single observation per treatment period, i.e. one time point before and after the tax implementation, and thus the only relevant time dimension spans across different treatment periods, which we refer to as $t$-time. We will then extend this method to the setting with multiple observations per treatment period in Section~\ref{sec:methods:multiple}. 

For each unit, we observe a baseline covariate vector, $\mathbf{X}_i$, and a binary treatment group indicator, $A_i$. Then, let $Z_{it} = tA_i$ represent the treatment status of unit $i$ in period $t$. We label the treatment and treatment status vectors for the entire population as $\mathbf{A}$ and $\mathbf{Z_t}$. Finally, we denote the outcome in each period for each unit as $Y_{it}$.

Each unit has a potential outcome in each treatment period under each population treatment assignment, $Y_{it}^{(\mathbf{Z_t})}$, resulting in $2^n$ potential outcomes per unit per period. This number is typically reduced to $2$ by invoking SUTVA, which mandates that the potential outcome of a particular unit is unaffected by the treatment status of other units, i.e. that $Y_{it}^{(\mathbf{Z_t})} = Y_{it}^{(Z_{it})}$. However, the presence of cross-border shopping violates this assumption in our setting, as stores would seemingly have different sales depending on the tax policy of nearby stores.

To address this concern, \cite{Aronow2013} introduced a modified SUTVA that reduces the number of potential outcomes while still accounting for the presence of interference. Their framework assumes that the population treatment status can only affect the potential outcome of unit $i$ at time $t$ through the unit's own treatment status at $t$, $Z_{it}$, and some known scalar function, $h_{it}: \{0, 1\}^n \rightarrow \mathbb{R}$. This function, referred to as the exposure mapping, represents the exposure level received by a unit that is not directly through their own treatment status.  Letting $g_{it}(\mathbf{Z_t}) = (Z_{it}, h_{it}(\mathbf{Z_t}))^T$ represent the exposure status of unit $i$ at time $t$ and invoking this modified SUTVA, we can then write the potential outcomes for any possible treatment assignment vector, $\mathbf{z_t}$, as: \setassumptioncounter 
\begin{equation}
\label{assump:mod_sutva}
Y_{it}^{(\mathbf{z_t})} = Y_{it}^{(g_{it}(\mathbf{z_t}))}
\end{equation}
% The assumption that treatment has no causal effect before its implementation, often formalized as a \textit{no anticipation} assumption is implicit in this formulation. 

\setequationcounter

In our study, we assume that the sales of store $i$ at time $t$ only depend on the tax policy through the store’s own tax status and the presence of a nearby store with a different tax status. Specifically,  we assume that  (\ref{assump:mod_sutva}) holds under:
\begin{equation}
\label{eqn:mod_sutva}
h_{it}(\mathbf{Z_t})=
\begin{cases}
1 & \text{if }  Z_{it}=0 \text{ and adjacent to any taxed region}\\
0 & \text{if }  Z_{it}=0 \text{ and not adjacent to any taxed region}\\
0 & \text{if }  Z_{it}=1 \text{ and adjacent to any untaxed region}\\
1 & \text{if }  Z_{it}=1 \text{ and not adjacent to any untaxed region}\\
\end{cases}
\end{equation}
This exposure mapping reduces the number of potential outcomes per store from $2^n$ to $4$ while still allowing for sweetened beverage sales in a given store to be affected by cross-border shopping to or away from neighboring regions. In the post-tax period of our study, we observe the control exposure status, $g_{i1}(\mathbf{Z_1})=(0,0)$, for Baltimore and other PA county stores not adjacent to Philadelphia (Non-Border); the treated exposure status, $g_{i1}(\mathbf{Z_1})=(1,0)$, for Philadelphia stores; and the neighboring control exposure status, $g_{i1}(\mathbf{Z_1})=(0,1)$, for other PA county stores adjacent to Philadelphia (Border).

\setequationcounter

\subsection{Policy-Relevant Causal Estimands}
\label{sec:methods:estimands}

Our goal here is to define causal estimands representing two policy-relevant questions. We first ask, what would be the average difference between sweetened beverage sales for Philadelphia stores in 2017 with and without the implemented PBT? This question corresponds to the ATT and can be defined in terms of potential outcomes as:
\begin{equation}
\label{eqn:att}
ATT := E[Y_1^{(1,0)} - Y_1^{(0,0)}|g(\mathbf{A})=(1,0)] = E[Y_1^{(1,0)} - Y_1^{(0,0)}|A=1]
\end{equation}
where we drop the unit-specific subscript, $i$, for convenience and define $g_i(\mathbf{A})$ analogously to $g_{it}(\mathbf{Z_t})$. The second equality holds by noting that no units in our study have $g_{i}(\mathbf{A})=(1,1)$.

We additionally ask, what would be the average difference between sweetened beverage sales for Border stores in 2017 with and without the implemented PBT? This question corresponds to what we call the \textbf{A}verage \textbf{T}reatment Effect on the \textbf{N}eighboring Control (ATN) and can be defined in terms of potential outcomes as:
\begin{equation}
\label{eqn:atn}
ATN:= E[Y_1^{(0,1)} - Y_1^{(0,0)}|g(\mathbf{A})=(0,1)]
\end{equation}

\subsection{Identifiability under the DiD Framework}
\label{sec:methods:did}

Since we do not observe the post-tax potential outcomes for unit $i$ under all possible combinations of $g_i(\mathbf{A})$, we cannot directly identify the ATT and ATN without further assumptions. Under the DiD framework and our proposed exposure mapping, we require the aforementioned counterfactual parallel trends assumptions between both the treated and control exposure groups to identify (\ref{eqn:att}): 
\setassumptioncounter
\begin{equation}
    \label{assump:att_parallel}
    E[Y_{1}^{(0,0)} - Y_{0}^{(0,0)}|A_i=1,\mathbf{X}] = E[Y_{1}^{(0,0)} - Y_{0}^{(0,0)}|g(\mathbf{A})=(0,0),\mathbf{X}] 
\end{equation} 
as well as between the neighboring control and control exposure groups to identify (\ref{eqn:atn}): 
\refstepcounter{assumptioncounter} \begin{equation}
% \setassumptioncounter 
\label{assump:atn_parallel}
    E[Y_{1}^{(0,0)} - Y_{0}^{(0,0)}|g(\mathbf{A})=(0,1),\mathbf{X}] = E[Y_{1}^{(0,0)} - Y_{0}^{(0,0)}|g(\mathbf{A})=(0,0),\mathbf{X}]
\end{equation}
whereby invoking the assumptions conditional on observed covariates, we allow for observed confounding often present in quasi-experimental settings. In our study, the counterfactual parallel trends assumptions would be violated if, for example, (i) wealthier populations are less likely to consume sweetened beverages in the post-tax period than the pre-tax period regardless of tax status, (ii) the distribution of wealth varies by region, \textbf{and} (iii) at least one of the following are true: (a) we do not observe this measure of wealth or (b) the distribution of wealth changes between tax periods. Otherwise, the assumptions would still hold. 

\setequationcounter

In addition to (\ref{assump:mod_sutva}), (\ref{assump:att_parallel}), and (\ref{assump:atn_parallel}), we also require the consistency and positivity assumptions, which respectively state that the observed outcome is equal to the potential outcome under the observed exposure status and that all units in the study have a non-zero probability of assignment to each of the exposures given the observed covariates.

\subsection{DiD Estimators for the ATT and ATN}
\label{sec:methods:estimators}

Here, we extend existing doubly robust estimators for the ATT under SUTVA in order to estimate both the ATT and ATN under our modified SUTVA with a binary exposure mapping, $h_{it}$. To do so, we denote $R_i$ as a binary indicator representing assignment to the exposure group of interest for the particular estimand and $Z^r_{it} = tR_i$ as an indicator representing the status of this relevant exposure at time $t$. Specifically, $R_i=1$ refers to the treatment group in the ATT comparison and the neighboring control group in the ATN comparison, whereas $R_i=0$ refers to the control group in both. We focus on methods for longitudinal panel data, corresponding to our data example. However,  these methods can be readily adapted to the setting where pre- and post-tax data are collected on different populations, i.e. cross-sectional data (\cite{Santanna2020}). 

\subsubsection{Two-Way Fixed Effects}

We begin by reviewing the commonly applied TWFE approach, which posits the linear outcome model: 
\begin{equation}
\label{eqn:twfe_unconditional}
    Y_{it} = \beta_0 + \mathbf{\beta}^T\mathbf{X_i} + \alpha_1 t + \alpha_2 R_i + \tau^{\text{fe}}Z^r_{it} + \epsilon_{it}
\end{equation}
where $\epsilon_{it}$ can be correlated within unit $i$. In the setting with two treatment periods, two exposure groups, and a single observation per treatment period, the coefficient $\tau^{\text{fe}}$ has been shown to be equivalent to the classical difference-in-means DiD estimator (\cite{Bertrand2004}) and thus can estimate the ATT or ATN under the unconditional analogues of (\ref{assump:att_parallel}) and (\ref{assump:atn_parallel}), where the expectations are not conditioned on $\mathbf{X}$.

However, the linear additive outcome model is limited by strict parametric assumptions. Further, numerous recent works have cited issues with its implicit assumption of a homogeneous treatment effect (\cite{DeChaisemartin2022}). We provide a simple example of the bias induced by this approach in the presence of time-varying confounding and heterogeneous treatment effects in Appendix~\ref{sec:appendix:twfe}.

\subsubsection{Outcome Regression Estimators}

Alternatively, one can attempt to impute the counterfactual outcome trends for the exposed group by modeling outcome dynamics under the control exposure. Here, we can apply the outcome regression (OR) estimator, first developed for the ATT in \cite{Heckman1997}, to estimate the ATT and ATN in our setting. The estimator plugs an estimate for $\mu_\Delta(\mathbf{X}) = E[Y_{1} - Y_{0}|g(\mathbf{A})=(0,0), \mathbf{X}]$ into:
\begin{equation}
    \label{eqn:or}
    \hat{\tau}^{or} = \mathbb{E}_{n_r}[(Y_{1}-Y_{0})-\hat{\mu}_\Delta(\mathbf{X})]
\end{equation}
where $\mathbb{E}_{n_r}$ denotes the empirical mean over the exposed group population. In contrast to (\ref{eqn:twfe_unconditional}), $\hat{\mu}_\Delta$ is generally estimated using only control group data to avoid specifying treatment effect dynamics and can be estimated with more flexible models. Still, this approach relies entirely on the correct specification of a model relating covariates to complex outcome dynamics.

\subsubsection{Inverse Probability Weighting Estimators}

Instead of modeling outcome dynamics, one can use a weighted estimator to balance confounding between the exposed and control groups. Here, we can apply the semi-parametric IPW estimator for the ATT by \cite{Abadie2005} to estimate the ATT and ATN in our setting. The estimator relies on the propensity score, or probability of assignment to the exposure of interest, $\pi_r(\mathbf{X})=P(R=1|\mathbf{X})$. In the case of panel data, the weights are calculated as:
\begin{equation}
    \label{eqn:weights}
    w_i = \frac{R_i - \pi_r(\mathbf{X_i})}{P(R_i=1)(1-\pi_r(\mathbf{X_i}))}
\end{equation}
and are used to estimate the causal effect after estimating $\pi_r(\mathbf{X})$ as:
\begin{equation}
    \label{eqn:ipw}
    \hat{\tau}^{ipw} = \mathbb{E}_n[\hat{w} (Y_{1}-Y_{0})]
\end{equation}
where the empirical mean is now taken over the study population of both the exposed and control groups. While flexible, IPW approaches can be unstable in finite samples or cases of nonoverlap when the propensity score is close to one for certain units.

\subsubsection{Doubly Robust Estimators}

Rather than choosing between the OR and IPW approaches, we propose applying the influence function (IF)-based DiD estimators developed for the ATT by \cite{Li2019} and \cite{Santanna2020} to estimate both the ATT and ATN under our binary exposure mapping. In addition to doubly robust (DR) properties, these estimators are also asymptotically normal and approach the semi-parametric efficiency bound when both nuisance functions are well-specified. For a deeper technical discussion as well as proofs of the properties of these estimators, we point the reader to \cite{Santanna2020}. 

The doubly robust (DR) plug-in estimator then incorporates estimates for both the propensity score and the outcome trend under no treatment to estimate the causal effect using:
\begin{equation}
    \hat{\tau}^{dr} = \mathbb{E}_n[\hat{w} ((Y_{1}-Y_{0})-\hat{\mu}_\Delta(\mathbf{X}))]
    \label{eqn:dr_formula}
\end{equation}
where $w$ and $\mu_\Delta(\mathbf{X})$ are defined as in the preceding estimators. As opposed to the TWFE approach, the IPW, OR, and DR approaches can easily incorporate non-parametric estimation of these nuisance functions using machine learning.

\subsection{Extension to Multiple Observations Setting}
\label{sec:methods:multiple}

Until now, we have presented methods for data with a single observation per treatment period. However, as noted earlier, our study comprises 13 different time points in both the pre- and post-tax periods. In this section, we present a simple yet robust approach for extending these methods to the multiple time point setting. The proposed approach turns out to be  a special case of the general framework proposed by \cite{Callaway2021} when there is no variation in treatment initiation time.

Let $m=1,...,n_m$ index the observation times in the \textit{post-tax} period. We then refer to the time dimension within a treatment period as $m$-time. Denoting $m$-time specific observations by adding a $m$-subscript to our previous notation, we can define $m$-time specific effects as:
\begin{equation}
\label{eqn:m_effects}
ATT(m) := E[Y^{(1,0)}_{1,m} - Y^{(0,0)}_{1,m}|A=1] \text{, } \hspace{3mm} ATN(m):= E[Y^{(0,1)}_{1,m} - Y^{(0,0)}_{1,m}|g(\mathbf{A})=(0,1)]
\end{equation}
We can then average across these $m$-time specific effects to summarize the entire effect: 
\begin{equation}
\label{eqn:effects_agg}
ATT= \frac{1}{n_m} \sum\limits_{m=1}^{n_m} ATT(m),\text{ } 
ATN= \frac{1}{n_m} \sum\limits_{m=1}^{n_m} ATN(m)
\end{equation}
% Identifying (\ref{eqn:effects_m_att}) and (\ref{eqn:effects_m_atn}) requires time-specific conditional counterfactual parallel trends assumptions. 
In our study, observations are observed at 4-week intervals and occur at the same calendar time in the pre- (2016) and post-tax (2017) periods. Thus, we consider data at each of these $n_m$ pairs as a two treatment period, single observation per period comparison and therefore require conditional counterfactual parallel trends assumptions at each $m$-time to identify (\ref{eqn:m_effects}), and thus (\ref{eqn:effects_agg}), as causal effects. Finally, we specify our nuisance functions across $m$-time as $\mu_{\Delta,m}=E[Y_{1,m} - Y_{0,m}|g(\mathbf{A})=(0,0), \mathbf{X}]$ and $\pi_{r,m} = P(R_m=1|\mathbf{X})$.

\section[]{Philadelphia Beverage Tax Analysis}
\label{sec:rda}

\begin{table}
 \vspace*{-6pt}
 \centering
 \bgroup
\def\arraystretch{1.25}%
 \def\~{\hphantom{0}}
  \begin{minipage}{150mm}
  \caption{Descriptive statistics for Philadelphia beverage tax study. Mean (standard deviation) metrics are calculated from 2016 data across stores in a given subset. Price of sweetened beverages is first averaged across 2016 time periods per store.}
  \label{tab:tab1}
  \begin{tabular*}{\textwidth}{@{}r@{\extracolsep{\fill}}c@{\extracolsep{\fill}}c@{\extracolsep{\fill}}c@{\extracolsep{\fill}}c}
  \toprule
  & Philadelphia & Baltimore & Border & Non-Border\\[1pt]
 \multicolumn{1}{l}{\textit{SGMs}} & \textit{(n=40)} & \textit{(n=15)} & \textit{(n=19)} & \textit{(n=51)}\\
 \midrule
  Price ($\$$/ oz)  & 5.85 (0.62) & 5.56 (0.47) & 5.41 (0.52) & 5.76 (0.81) \\
 White ($\%$) & 54.9 (29.5) & 42.2 (27.3) & 81.8 (15.2) & 89.2 (6.1) \\
 Black ($\%$) & 34.2 (31.3) & 52.8 (30.1) & 10.3 (15.5) & 5.1 (3.8) \\
 Income ($\$1000$) & 43.8 (12.4) & 51.2 (13.3) & 79.9 (19.4) & 81.3 (20.0)\\
 House Value ($\$1000$) & 173.3 (76.6) & 174.3 (57.5) & 318.4 (86.5) & 298.8 (91.8) \\
 Mass Merchandiser ($\%$) & 35.0 & 13.3 & 21.1 & 33.3 \\
 \midrule
 \multicolumn{1}{l}{\textit{Pharmacies}} & \textit{(n=140)} & \textit{(n=45)} & \textit{(n=32)} & \textit{(n=78)}\\
 \midrule
 Price ($\$$/ oz) & 7.52 (0.58) & 6.85 (0.59) & 7.00 (0.58) & 7.55 (0.49)\\
 White ($\%$) & 48.1 (28.9) & 36.4 (24.7) & 69.1 (24.9) & 89.7 (5.5) \\
 Black ($\%$) & 40.1 (31.2) & 59.3 (26.9) & 23.5 (25.2) & 4.9 (3.4)\\
 Income ($\$1000$) & 43.6 (16.2) & 44.9 (15.5) & 73.2 (21.3) & 81.9 (18.5) \\
 House Value ($\$1000$) & 182.4 (103.3) & 174.9 (60.2) & 272.3 (104.9) & 308.8 (82.5) \\
\hline
\end{tabular*}\vskip18pt
\end{minipage}
\egroup
 \vspace*{6pt}
\end{table}

% \begin{minipage}{\linewidth}% to keep image and caption on one page
% \makebox[\linewidth]{%        to center the image
%   \includegraphics[keepaspectratio=true,scale=0.6]{slike/visina8}}
% \captionof{figure}{...}\label{visina8}%      only if needed  
% \end{minipage}
\begin{minipage}{\linewidth}% to keep image and caption on one page
\makebox[\linewidth]{%
 \includegraphics[width=0.6\columnwidth]{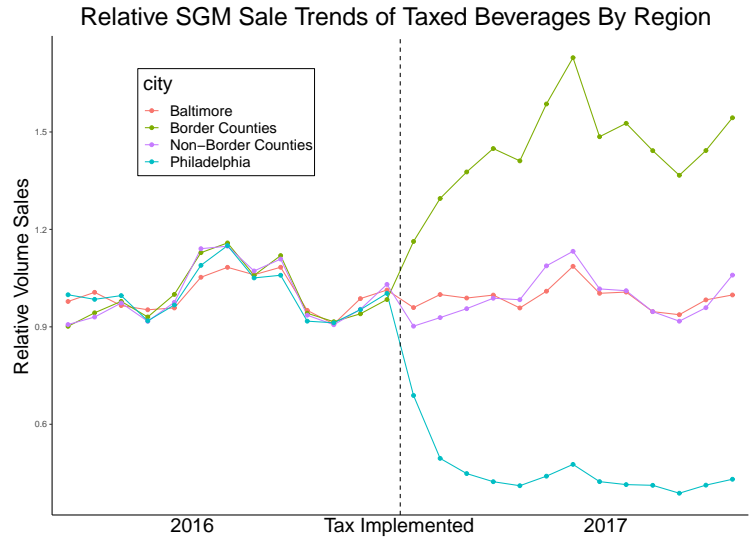}}
%  \centerline{\includegraphics[width=\columnwidth]{Figures/relative_pharmacy_trends.eps}}
% \caption{Regional average of taxed beverage sales per store taken relative to average 2016 sales. Top: sales for supermarkets, grocery stores, and mass merchandisers. Bottom: sales for pharmacies.}
\captionof{figure}{Regional average of taxed beverage sales per store taken relative to average 2016 sales for supermarkets, grocery stores, and mass merchandisers.}
\label{fig:salestrends}
\end{minipage}

\subsection{Descriptive Analyses}

We started by grouping stores by region (Philadelphia, Baltimore, Border, Non-Border) and type (SGM, pharmacy). We analyzed SGMs and pharmacies separately due to expected differences in sales dynamics that may be difficult to properly model (\cite{Roberto2019}). Pre-treatment covariates for our study are summarized in Table~\ref{tab:tab1}. Notably, regional-level covariates are similar between Philadelphia and Baltimore as well as between Border and Non-Border regions, but not between Philadelphia and Non-Border or Border and Baltimore regions. Therefore, we used Baltimore stores as the control group for Philadelphia stores in the ATT comparison and Non-Border stores as the control group for Border stores in the ATN comparison.

Regions affected by the PBT either directly or through bypass effects demonstrate a clear disruption in volume sales of sweetened beverages between the year before and after tax implementation relative to our control regions, as visualized in Figure~\ref{fig:salestrends}. The sales in Philadelphia SGMs and pharmacies respectively decreased by $54.90\%$ and $21.57\%$ from 2016 to 2017, whereas neighboring SGMs and pharmacies increased by $44.75\%$ and $21.60\%$. This comes in stark contrast to the relatively constant SGM beverage sales in Baltimore ($0.96\%$ decrease) and Non-Border stores ($0.86\%$ decrease) and slightly decreasing pharmacy beverage sales ($10.01\%$ Baltimore, $11.74\%$ Non-Border).

\begin{figure}[H]
\centering
 % \def\~{\hphantom{0}}
% \begin{minipage}{165mm}
     \includegraphics[width=0.5\textwidth]{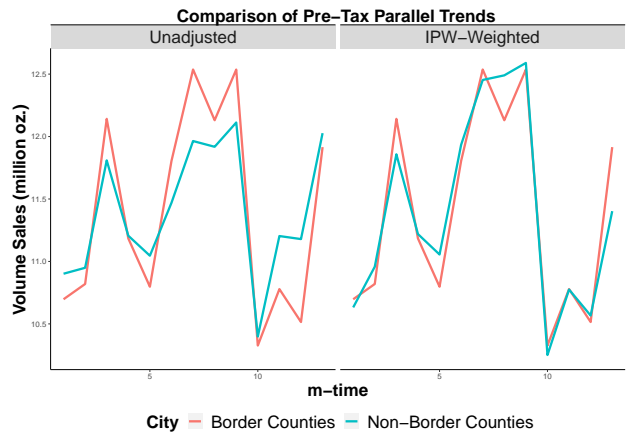}
     \caption{An example of how conditioning on confounders can affect estimation. IPW-weighting makes pre-tax parallel trends more plausible between Border and Non-Border county pharmacies.}
    %  Right: DR methods estimate a larger drop in Philadelphia pharmacy beverage sales due to the tax at early ($m=1,...,7$) time points.}
     \label{fig:conditional_justification}
     % \end{minipage}
\end{figure}

\subsection{Estimation of Treatment Effects}

\begin{table}
 \centering
 \def\~{\hphantom{0}}
 \begin{minipage}{165mm}
  \caption{Beverage Tax Effect Estimates}
\label{tab:estimates}
  \begin{tabular*}{\textwidth}{@{}r@{\extracolsep{\fill}}c@{\extracolsep{\fill}}c@{\extracolsep{\fill}}c@{\extracolsep{\fill}}c}
  \toprule
 & \multicolumn{2}{c}{{ATT}} & \multicolumn{2}{c}{{ATN}}  \\ [1pt]
\cline{2-3} \cline{4-5}  \\ [-6pt]
 & TWFE & DR & TWFE & DR \\
 \midrule
 \multicolumn{1}{l}{\textit{SGMs (\textit{million oz.})}} & & & & \\
 Winter & -1.90 & -1.83 & 0.88 & 0.87 \\ 
 & (-2.44, -1.41) & (-2.50, -1.14) & (0.56, 1.26) & (0.55, 1.23) \\ 
 Spring & -2.41 & -2.57 & 1.14 & 1.13 \\ 
 & (-3.10, -1.74) & (-3.47, -1.86) & (0.66, 1.71) & (0.68, 1.66) \\ 
 Summer & -2.64 & -2.50 & 1.29 & 1.29 \\ 
 & (-3.47, -1.87) & (-3.39, -1.71) & (0.77, 1.89) & (0.79, 1.89) \\ 
 Fall & -2.28 & -2.30 & 1.23 & 1.23 \\ 
 & (-3.00, -1.60) & (-3.11, -1.55) & (0.72, 1.78) & (0.78, 1.77) \\ 
 Annual & -2.30 & -2.30 & 1.14 & 1.14 \\ 
 & (-2.99, -1.66) & (-2.98, -1.64) & (0.68, 1.66) & (0.72, 1.64) \\
 \multicolumn{1}{l}{\textit{Pharmacies (\textit{thousand oz.})}} & & & & \\
 Winter & -5.7 & -14.1 & 29.3 & 35.4 \\ 
 & (-16.1, 5.0) & (-26.8, 0.1) & (16.3, 46.4) & (17.4, 55.5) \\ 
 Spring & -24.6 & -27.8 & 34.3 & 40.8 \\ 
 & (-36.0, -12.1) & (-39.0, -15.0) & (20.4, 51.9) & (24.0, 61.2) \\ 
 Summer & -34.1 & -36.3 & 32.5 & 40.5 \\ 
 & (-46.5, -21.6) & (-48.0, -24.5) & (18.6, 51.6) & (22.2, 66.2) \\ 
 Fall & -18.3 & -18.9 & 35.1 & 41.0 \\ 
 & (-30.1, -6.4) & (-31.5, -5.5) & (21.2, 54.3) & (23.5, 64.8) \\ 
 Annual & -20.5 & -23.9 & 33.0 & 39.5 \\ 
 & (-30.3, -10.3) & (-34.4, -12.1) & (19.8, 50.9) & (23.1, 61.1) \\ 
\bottomrule
\end{tabular*}
\end{minipage}
{\footnotesize Seasons are defined chronologically, with Winter as the first 3 observations of the calendar year, Spring as the next 3, Summer as the following 3, and Fall as the final 4.}
\end{table}

We estimated the ATT and ATN using both the standard TWFE approach, which requires the standard unconditional counterfactual parallel trends, and our proposed DR approach, which requires conditional counterfactual parallel trends. To estimate the ATT and ATN using the TWFE approach, we estimated time-specific treatment effects using the linear model in (\ref{eqn:twfe_unconditional}) for each $m=1,...,13$. To estimate the ATT and ATN using our proposed doubly robust methods, we fit a different linear regression model for each $\mu_{\Delta,m}$ and logistic regression model for each $\pi_{r,m}$, with terms for our observed covariates (\cite{Stuart2014}).  

To estimate $95\%$ confidence intervals (CIs) for our effect estimates, we implemented a stratified nonparametric bootstrap sampling approach (\cite{EfronTibshirani1993}). For each of the four regions, we re-sampled with replacement from the empirical distribution of the regional subsample, where a store’s entire observed data vector is re-sampled. Stratifying by region limits extreme bootstrap samples where certain regions may only have a few representative members, which is the case in our study. We then estimated the $2.5$ and $97.5$ percentiles among the 500 bootstrap replicates as our interval bounds. A brief discussion and comparison of possible CI approaches is presented in Appendix~\ref{sec:appendix:ci}.

To bolster the credibility of Assumption~(\ref{assump:mod_sutva}) under (\ref{eqn:mod_sutva}), we removed Non-Border stores in PA zip codes within 6 miles of the Philadelphia border (138 stores). We made this decision after estimating a small but nonzero ATN on these stores using our DR methodology in a preliminary analysis (Table~\ref{tab:exposure_mapping}), suggesting this group may contain a mixture of stores with control and neighboring control exposures. Assumption~(\ref{assump:mod_sutva}) further implies that the treatment has no causal effect before its implementation, often formalized as a \textit{no anticipation} assumption, which would be violated in our study if Philadelphia residents stock-piled sweetened beverages in the months leading up to the tax. To evaluate this assumption, we used our DR methodology to estimate the ATT on SGMs and pharmacies between the first pre-tax observation time and the $m^{\text{th}}$ pre-treatment observation time, $ATT^{(pre)}(m) := E[Y^{(1,0)}_{0,m} - Y^{(0,0)}_{0,1} | A=1]$, for $m=12,13$ (Table~\ref{tab:pretrends}). Our $95\%$ CIs included the null effect except for the $ATT^{(pre)}(12)$ of pharmacies, which was statistically significant but negative suggesting, if anything, that consumers were pre-emptively cross-border shopping rather than stock-piling.

Researchers commonly use tests on pre-treatment parallel trends to assess the plausibility of the counterfactual parallel trends assumption in DiD studies (\cite{Bilinski2018NothingAssumptions}). In our setting these tests can be conducted robustly to assess Assumptions (\ref{assump:att_parallel}) and (\ref{assump:atn_parallel}) by using our proposed methodology to estimate $ATT^{(pre)}(m)$ and the analogous $ATN^{(pre)}(m)$ for $m=2,\dots,13$. If the pre-treatment outcome trends are parallel between the exposure groups, these effects would be zero. As \cite{Bilinski2018NothingAssumptions} note, practitioners should be wary to conduct these tests under the null hypothesis of parallel trends as this would reward highly uncertain tests that fail to rule out large violations of parallel trends. Therefore, we report $95\%$ CIs of such tests as evidence to reject violations of parallel trends outside the estimated bounds. These intervals, provided in Table~\ref{tab:pretrends}, largely include $0$. A visual example of how conditional parallel pre-trends may be more plausible than the unconditional analogy is given in Figure~\ref{fig:conditional_justification}. Still, these tests are a limited proxy for the required counterfactual parallel trends and rely on the assumption that parallel trends within the pre-treatment period can be extrapolated to counterfactual trends between tax periods.

ATT and ATN estimates for sweetened beverage sales at SGMs and pharmacies aggregated by season and year  appear in Table \ref{tab:estimates}. In the year after tax initiation, SGMs see an average loss of 2.26 million oz. (95\% CI: (1.54, 2.95)) in Philadelphia per 4-week period and gain of 1.20 million oz. (95\% CI: (0.79, 1.70)) in neighboring stores. Pharmacy stores see an average loss of 22.9 thousand oz. (95\% CI: (11.2, 32.5)) in Philadelphia per 4-week period and a 38.3 thousand oz. (95\% CI: (23.6, 56.9)) gain in neighboring stores. 

Interestingly, the ATN is larger in magnitude than the ATT for pharmacies. Since our analyses are at the store-level, we would need to consider the total number of stores in each region to assess the relative magnitudes at the regional level. Additionally, some Philadelphia consumers may respond to the PBT by switching their sweetened beverage purchases from Philadelphia SGMs to Border county pharmacies.

Our results show considerable effect heterogeneity between seasons for both the ATT and ATN, with effects strongest in warmer seasons. ATT effects are strongest in the Summer for SGMs and Pharmacies, with estimated decreases of 2.47 million oz. (95\% CI: (1.64, 3.29)) and 34.8 thousand oz. (95\% CI: (20.8, 46.1)) due to the tax, respectively. ATN effects are strongest for SGMs in the Summer (increase of 1.35 million oz., 95\% CI: (0.86, 1.90)) but the largest effects are estimated for pharmacies in the Fall (increase of 42.0 thousand oz., 95\% CI: (26.3, 60.9)). Related, ATT effects are weakest in the winter with decreases of 1.93 million oz. for SGM (95\% CI: (1.37, 2.52)) and 12.7 thousand oz. for pharmacies (95\% CI: (0.7, 23.5)). This holds true for the ATN of SGMs (increase of 0.90 million oz., 95\% CI: (0.57, 1.27)) and pharmacies (increase of 35.4 thousand oz., 95\% CI: (20.5, 53.2)). Such heterogeneity matches our expectations as the warmer temperatures and general increase of sweetened beverage sales in warmer months may further incentivize consumers to travel to bypass the PBT. However, the subdued tax effects in Winter may also be indicative of a gradual consumer response to the PBT. 

The DR estimates closely resemble the estimates using the standard TWFE methodology for SGMs. This may indicate some plausibility of the unconditional parallel trends assumption between exposure groups for these stores but also reflects that our available covariates were not very informative of the outcome for this set of stores. For pharmacies, the DR methods produce higher magnitude effect estimates than their TWFE counterparts, which may suggest that confounding is masking some of the tax effect in standard analyses. For example, our outcome models for $\mu_{\Delta,m}$ associate lower percentages of White residents with higher declines in beverage sales at Baltimore pharmacies between 2016 and 2017. Since Philadelphia pharmacies are in zip codes with higher percentages of White residents, we would underestimate the post-tax counterfactual sweetened beverage sales of Philadelphia stores by not properly adjusting for race. Notably, by accounting for confounders the DR CIs for the Winter ATT on pharmacies do not include zero unlike the TWFE CIs.

\subsection{Estimation of Effects by Geographical Proximity}

While population-level effects are helpful, policies may affect subpopulations within each region differentially. Understanding \textit{who} policies are affecting most is especially helpful for policy makers when deciding whether to continue policies, how to address disparities induced by policies, and how to implement policies in regions with different population compositions. 

To help understand how policy effects may vary by geographic proximity to non-taxed regions, we first defined subgroups of Philadelphia zip codes according to their border status – PA-bordering, New Jersey (NJ)-bordering, and Non-Bordering – which are visualized in Figure~\ref{fig:clustermap_border} in yellow, purple, and orange, respectively. We then used our proposed DR methodology to estimate an annual \textit{relative} sales effect, $ATT_{rs} := E[Y_1^{(1,0)}|g(\mathbf{A})=(1,0)] / E[Y_1^{(0,0)}|g(\mathbf{A})=(1,0)]$ for pharmacies in each of these subgroups. The sales ratio provides a comparable scale for subpopulations that may differ in magnitude of sweetened beverage sales, with more discussion presented in Appendix~\ref{sec:appendix:mult_effect}. The subgroups contained 58, 26, and 56 pharmacies with estimated effects of -33\% (95\% CI: (-39\%, -26\%)), -10\% (-24\%, +5\%), and -6\% (-17\%, +6\%), respectively, suggesting that PA-bordering Philadelphia pharmacies experienced a larger decrease than those adjacent only to NJ or other Philadelphia zip codes, for which bypass may be less practical (e.g., requires a toll to enter). These findings complement previous studies which found reduced tax effects on Philadelphia residents further from the city border (\cite{Cawley2019}). 

\begin{minipage}{\linewidth}
\makebox[\linewidth]{
    \centering
    \includegraphics[width=0.8\columnwidth]{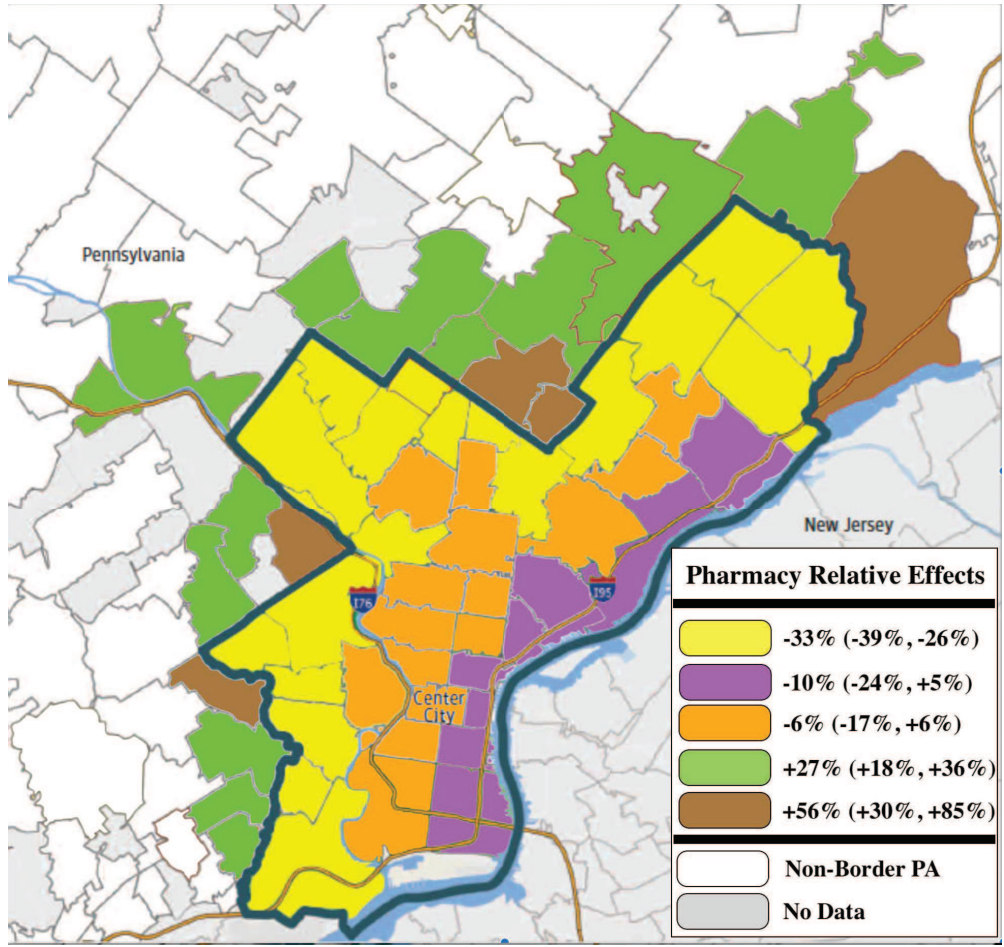}}
    \captionof{figure}{Clustering zip codes by geographic factors (border status for Philadelphia stores and proxies for available traffic and sales from proximal taxed stores for Border county stores) suggests heterogeneity in tax effects.}
    \label{fig:clustermap_border}
\end{minipage}

To further understand the influence of geographic heterogeneity on bypass effects, we also defined subgroups of Border county zip codes according to measures of proximal taxed population and the year-to-year (YtY) differences in taxed beverage sales of these populations. Specifically, we took the total population and YtY difference of each PA-bordering Philadelphia zip code and divided these measures by the number of Border county zip codes they were adjacent to, as proxies for the amount of traffic and sales ``available'' from a taxed zip code. For each Border county zip code, we then took the sum of these measures from all adjacent Philadelphia zip codes as a proxy for the amount of available traffic and sales from proximal taxed stores. We used K-means clustering to assign these zip codes to groups with low and high amounts of available traffic and sales. Using our proposed method, we estimated substantially lower effects in Border county zip codes with low available traffic and sales from taxed stores (27\% increase, 95\% CI: (18\%, 36\%)) than those with high available traffic and sales from taxed stores (56\% increase, 95\% CI: (30\%, 85\%)), although the CIs are quite wide in the latter group which is likely due to the small subgroup size (10 pharmacies) and the large heterogeneity of our proxy measures within this subgroup. While these exploratory analyses demonstrate potential geographic heterogeneity in policy effects, we cannot differentiate between effect heterogeneity due to geographical factors and heterogeneity due to other population dynamics associated with geography by estimating causal effects on different subgroup populations.

\section{Simulation Studies}
\label{sec:sims}

\subsection{Design}

We performed simulation studies motivated by the PBT study to evaluate the performance of different estimators under realistic scenarios. To generate our samples, we first simulate $\mathbf{X_i^{(orig)}} \sim N(\mathbf{0}, \mathbf{I_4})$ for each unit $i=1,\dots,n$. We coerce one of the covariates to vary across $m$-time by setting $X^{(orig)}_{i,4,1} = X^{(orig)}_{i,4}$ and simulating $X^{(orig)}_{i,4,m} \sim N(X^{(orig)}_{i,4,m-1} + 0.25, 0.25)$ for $m = 2,3,\dots,n_m$. Then, we apply the transformation from \cite{KangSchafer2007} on $\mathbf{X^{(orig)}}$ to get $\mathbf{X^{(obs)}}$:

$X_{i,1}^{(obs)} = (0.6 + X_{i,1}^{(orig)}  X_{i,3}^{(orig)} / 25)^3$, \hspace{5mm} $X_{i,2}^{(obs)} = 10 + \frac{ X_{i,2}^{(orig)} } { 1 + \exp( X_{i,3}^{(orig)} )}$

$X_{i,3}^{(obs)} = \exp(0.5  X_{i,3}^{(orig)})$, \hspace{23mm} $X_{i,4,m}^{(obs)} = (20 + X_{i,2}^{(orig)} * X_{i,4,m}^{(orig)})^2$

In each simulation, our estimators use the observed covariates, $\mathbf{X^{(obs)}}$. However, the covariate set used to generate the exposure, $\mathbf{X^{(a)}}$, and outcome, $\mathbf{X^{(\mu)}}$, vary depending on the scenario. Units are split between the ATT and ATN comparisons according to a 43:57 ratio to mimic some of the imbalance seen in our dataset. Exposure within each comparison is then simulated according to binomial models with $P(g_i(\mathbf{A}) = (1,0) | \mathbf{X_i}) = expit(\mathbf{\beta^{(T)}}'\mathbf{X_i^{(a)}})$ and $P(g_i(\mathbf{A} | \mathbf{X_i}) = (0,1)) = expit(\mathbf{\beta^{(N)}}'\mathbf{X_i^{(a)}})$. For units with $g_i(\mathbf{A}) = (0,1)$, we additionally simulate a variable representing distance to the Philadelphia border, $D_i \sim N(1/(1+exp(X^{(obs)}_{i,2} + X^{(obs)}_{i,3} + X^{(obs)}_{i,4,1})), 0.05)$, and bound it between 0.1 and 0.9. This variable is used to induce treatment effect heterogeneity in the ATN. 

Outcomes are generated with the linear model as: $Y_{itm} = 20 + \alpha_i + \alpha_{g_i(\mathbf{A})} + \gamma_m + \gamma_t + t \{ \mathbf{\lambda_{tm}}'\mathbf{X_i^{(\mu)}} + \tau_m^{(ATT)}  \mathbbm{1}(g_i(\mathbf{A})=(1,0))  + \tau_m^{(ATN,*)}  \mathbbm{1}(g_i(\mathbf{A})=(0,1)) (1 - D_i)  \} + \epsilon_{itm}$. Here, $\alpha_i$ correspond to unit-specific random intercepts, $\alpha_{g_i(\mathbf{A})} \sim N(0,1)$ is a fixed effect for the exposure group, $\gamma_m$ and $\gamma_t$ are fixed effects per observation time and treatment period, $\mathbbm{1}$ is an indicator function, and $\epsilon_{itm} \sim N(0,0.5)$ are iid error terms. The ATT for the entire study period is given as the average of the specified parameters $\tau_m^{(ATT)}$ across $m=1,\dots,n_m$. The population-level ATN varies according to $D_i$ and is given by averaging $\tau_m^{(ATN)} = \frac{1}{\sum\limits_{i=1}^n \mathbbm{1}(g_i(\mathbf{A})=(0,1))} \sum\limits_{i=1}^n  \mathbbm{1}(g_i(\mathbf{A})=(0,1)) \tau_m^{(ATN,*)} (1-D_i)$ over $m=1,\dots, n_m$. Notably, our framework allows for heterogeneous confounding ($\mathbf{\lambda_{tm}}$) and treatment effects ($\tau_{m}$) over $m$-time, as well as heterogeneous treatment effects over confounders, like distance in this setting. Parameter settings used for the simulations can be found in Appendix~\ref{sec:appendix:sim_param}.

Simulations were generated according to three different combinations of sample size  ($n$) and number observation times ($n_m$) – (1) $n=250, n_m=13$, (2) $n=2000, n_m=2$ and (3) $n=500, n_m=4$. We consider four different scenarios depending on  the covariate set used to specify the treatment ($\mathbf{X^{(a)}}$) and outcome ($\mathbf{X^{(\mu)}}$) models – (a) $\mathbf{X^{(a)}}= \mathbf{X^{(obs)}}$, $\mathbf{X^{(\mu)}}= \mathbf{X^{(obs)}}$, (b) $\mathbf{X^{(a)}} = \mathbf{X^{(orig)}}$, $\mathbf{X^{(\mu)}} = \mathbf{X^{(obs)}}$, (c) $\mathbf{X^{(a)}} = \mathbf{X^{(obs)}}$, $\mathbf{X^{(\mu)}} = \mathbf{X^{(orig)}}$, and (d) $\mathbf{X^{(a)}} = \mathbf{X^{(orig)}}$, $\mathbf{X^{(\mu)}} = \mathbf{X^{(orig)}}$. Thus, our outcome model is correctly specified in (a) and (b) but misspecified in (c) and (d), whereas our propensity score model is correctly specified in (a) and (c) but misspecified in (b) and (d). 

In each scenario, we generate 1000 replicates to examine the performance of the described TWFE, IPW, OR, and DR methods. To estimate $\mu_{\Delta,m}$, we fit a linear regression model on the difference in outcomes, $\Delta Y_{im} = Y_{i1m} - Y_{i0m}$, for the control group using a separate model for $m=1,\dots,n_m$. To estimate $\pi_{r,m}$, we fit a single time-invariant logistic regression model. CIs are estimated with the aforementioned stratified bootstrap approach. 

\subsection{Results}

\begin{sidewaystable}
 \centering
 \def\~{\hphantom{0}}
\begin{minipage}{125mm}
  \hsize\textheight
  \caption{Simulation results}
\label{tab:sims}
\hskip-5pt\begin{tabular*}{\textheight}{@{}l@{\extracolsep{\fill}}c@{\extracolsep{\fill}}c@{\extracolsep{\fill}}c@{\extracolsep{\fill}}c@{\extracolsep{\fill}}c@{\extracolsep{\fill}}c@{\extracolsep{\fill}}c@{\extracolsep{\fill}}c@{\extracolsep{\fill}}c@{\extracolsep{\fill}}c@{\extracolsep{\fill}}c@{\extracolsep{\fill}}c@{\hskip5pt}}
  \toprule
 & \multicolumn{4}{c}{{Bias ($\%$)}} & \multicolumn{4}{c}{{Std. Err}} & \multicolumn{4}{c}{{Coverage ($\%$)}} \\ [1pt]
\cline{2-5} \cline{6-9} \cline{10-13} \\ [-6pt]
& & & & & & & & & & & & \\ [-3pt]
Scenario & TWFE & OR & IPW & DR & TWFE & OR & IPW & DR & TWFE & OR & IPW & DR
\\ 
 \midrule
 \multicolumn{1}{l}{\textit{ATT}} & & & & & & & & & \\
 \midrule
 1a &-12.784 & -0.069 & 0.929 & -0.103 & 0.564 & 0.045 & 0.389 & 0.047 & 91.9 & 94.1 & 98.5 & 95.0 \\ 
 1b &  -39.051 & -0.067 & -2.251 & -0.096 & 0.571 & 0.045 & 0.49 & 0.047 & 72.0 & 95.3 & 96.5 & 96.3 \\ 
 1c & -7.286 & -9.348 & -0.225 & -2.754 & 0.597 & 0.386 & 0.492 & 0.472 & 93.2 & 93.5 & 97.7 & 95.3 \\ 
 1d & 22.566 & 28.382 & 35.917 & 32.824 & 0.624 & 0.397 & 0.538 & 0.453 & 88.0 & 72.8 & 73.3 & 70.3  \\ 
 \midrule
 2a & -10.887 & -0.002 & -0.003 & 0.032 & 0.226 & 0.038 & 0.209 & 0.040 & 83.4 & 94.7 & 95.3 & 94.3 \\ 
 2b & -45.221 & -0.062 & -1.576 & -0.059 & 0.224 & 0.037 & 0.148 & 0.038 & 1.7 & 94.7 & 91.0 & 95.0  \\ 
 2c &  -5.278 & -8.913 & -0.391 & -0.745 & 0.208 & 0.138 & 0.199 & 0.187 & 90.6 & 75.2 & 94.6 & 93.7 \\ 
 2d & 19.201 & 29.893 & 35.902 & 33.978 & 0.206 & 0.134 & 0.171 & 0.141 & 35.8 & 0.6 & 1.1 & 0.0 \\
 \midrule
 \multicolumn{1}{l}{\textit{ATN}} & & & & & & & & & \\
 \midrule
 1a & -58.815 & 0.078 & 0.005 & 0.186 & 0.502 & 0.037 & 0.347 & 0.042 & 75.2 & 100.0 & 98.6 & 100.0 \\
 1b & -13.883 & -0.166 & -24.357 & -0.082 & 0.497 & 0.039 & 0.801 & 0.043 & 93.2 & 99.7 & 98.2 & 99.7 \\
 1c & 45.954 & -28.936 & -0.928 & -4.004 & 0.519 & 0.359 & 0.561 & 0.526 & 84.8 & 88.6 & 97.0 & 93.7 \\ 
 1d & 10.542 & -53.447 & -63.872 & -49.881 & 0.539 & 0.355 & 0.750 & 0.405 & 93.8 & 64.4 & 85.2 & 72.8 \\ 
 \midrule
 2a & -49.862 & -0.051 & 0.064 & 0.013 & 0.194 & 0.031 & 0.128 & 0.033 & 27.2 & 98.2 & 95.7 & 98.0 \\ 
 2b & -26.739 & 0.026 & -28.104 & 0.018 & 0.193 & 0.032 & 0.420 & 0.035 & 71.7 & 97.0 & 66.8 & 97.5 \\ 
 2c & 52.672 & -28.980 & 0.132 & -0.578 & 0.174 & 0.127 & 0.139 & 0.193 & 14.7 & 37.1 & 94.6 & 93.2 \\ 
 2d &  -5.510 & -51.219 & -58.066 & -42.967 & 0.175 & 0.120 & 0.233 & 0.200 & 84.8 & 0.4 & 2.7 & 11.8 \\ 
\bottomrule
\end{tabular*}
\end{minipage}
\end{sidewaystable}

Simulation results are summarized in Table~\ref{tab:sims}, with results from scenario 3 presented in Table~\ref{tab:sims_extra}. We evaluate each method according to the average bias and standard error of our point estimates as well as the coverage of our CIs. 

For both the ATT and ATN, the estimates using the standard TWFE approach are highly biased for all scenarios as the method does not account for time-varying confounding. The estimates using the OR and IPW approaches are unbiased in the scenarios where the respective model is correctly specified, whereas those from the DR approach are unbiased in scenarios (a)-(c). All approaches are biased in scenario (d) when both models are incorrectly specified. While the IPW approach appears relatively unbiased for the ATT in scenario (b) when the propensity score model is misspecified, we caution that this is a product of the specific data generating mechanism and note that this chance behavior is not expected to hold in general, as seen in the ATN comparison. However, the slight bias of the DR approach in scenario (c) in smaller sample settings is something that has been noted in previous works (\cite{Li2019, Santanna2020}), as the DR approach appears more dependent on the outcome model specification than that of the propensity score model. 

The OR and DR approaches result in the smallest standard errors, with the OR approach slightly more efficient in these finite sample settings but significantly less robust to misspecification. The higher efficiency of the outcome model when correctly specified has been noted by \cite{Li2019} and \cite{Santanna2020} as well. Notably, the IPW approach has relatively large standard errors even with a well-specified model, which is well-documented in the literature as a result of unstable weights in finite sample settings. 

Bootstrap approaches for CIs work well in these simulations, roughly achieving the nominal coverage probability when models are correctly specified. The intervals are slightly inflated in small sample settings (Scenario 1) and for the ATN. The latter observation may result from unmodeled treatment effect heterogeneity leading to efficiency loss. Still, when the models are correctly specified, the CIs tend to be conservative.

\section{Discussion}
\label{sec:disc}

Bypass effects often occur when a policy imposes restrictions on individuals, often substantiallly offsetting  the intended effects of the policy. Understanding these effects is  critical for policy makers and evaluators. In this work, we propose a framework to estimate policy-relevant causal effects in the presence of such interference by joining together the ideas of exposure mappings and doubly robust DiD estimators. We applied our methods to estimate the causal effects of the PBT on Philadelphia and its bordering counties, accounting for both interference and confounding. Notably, we estimated more pronounced effects of the tax on pharmacies than methods used in previous studies. Additionally, we have used our methods to reveal new insights concerning effect heterogeneity according to season and geographic proximity. 

It is important to note that we did not have access to sales data from NJ border stores, which may also see bypass effects, but perhaps to a lesser degree. Further, our estimates do not tell us what \textit{would} have happened in a counterfactual scenario where a tax was implemented in both Philadelphia and its surrounding counties, a situation that may be quite relevant to policy makers. In future analyses it would be of interest to  robustly account for residual spatial correlation between stores of the same region or auto-correlation between time-specific effect estimates. 
Finally, doubly robust methods that handle continuous and multi-dimensional exposure mappings instead of a known binary exposure mapping would be valuable for this study and many others. In addition to increasing efficiency, such methods would strengthen subpopulation analyses by allowing investigators to flexibly and efficiently understand how policy effects vary across space and/or other factors. In practice, these insights may shed light on why certain cities, such as Seattle, have seen less substantial bypass effects, which has been thought to be a product of geographical borders between neighboring counties (\cite{Powell2020}). However, collecting and incorporating accurate and relevant geographical and transportation (e.g. access to a car or public transportation) data  may still pose a challenge.

\section*{Acknowledgements}

This work was supported by NSF Grant 2149716 (PIs: Mitra and Lee).

\vspace*{-8pt}

%%%%%% include this section only if your manuscript refers to supplementary
%%%%%% materials -- see Instructions for Authors at 
%%%%%% http://www.tibs.org/biometrics

\section*{Available Code}

Code and an example simulated dataset are provided on GitHub at https://github.com/garyhettinger/DiD-interference.
\vspace*{-8pt}

\bibliographystyle{unsrtnat}
\bibliography{references}

\appendix

\label{sec:appendix}

\section{Concerns with TWFE approach in the Presence of Heterogeneous Treatment Effects}
\label{sec:appendix:twfe}

Many works have cited issues with the TWFE approach in studies where different groups within the population receive treatment at different times, i.e. staggered adoption, and the treatment effects are heterogeneous across these groups. While the concern of staggered treatment adoption does not apply to our setting, since all stores are exposed to the tax at the same point in time, it is noteworthy that the straightforward extension of the TWFE model to account for time-varying confounding:
\begin{equation}
\label{eqn:twfe_conditional}
Y_{it} = \beta_0 + \mathbf{\beta}^T\mathbf{X_i} + \mathbf{\theta}^T\mathbf{tX_i} + \alpha_1 t + \alpha_2 R_i + \tau^{\text{fe}}Z^r_{it} + \epsilon_{it}
\end{equation}
is not robust to treatment effects that are heterogeneous across $\mathbf{X}$. 

To demonstrate this limitation, we consider a simple simulation setting. We first generate $X_i \sim N(0,1)$. Then, we simulate $A_i \sim Bin(1/(1+exp(-0.5 + 0.5X_i))$. Finally, we simulate $Y_{it} \sim N(10 + t + 2A_i + 2X_i + t X_i + 4 t \theta_i, 0.1)$. In the setting with a homogeneous treatment effect, $\theta_i = 1$ for all $i$. In the setting with a heterogeneous treatment effect, we set $\theta_i = 1 + \mathbbm{1}\{X_1 \geq 0.5\}$. 

Then, the true ATT is 4 in settings with a homogeneous treatment effect but $4+P(X \geq 0.5 | A = 1)$ in settings with heterogeneous treatment effects. We run 1000 simulations with $n=10000$ and fit a linear model $Y_{it} = \beta_0 + \beta_t t + \beta_a A_i + \beta_x X_i + \beta_{xt} tX_i +\tau t A_i + \epsilon_{it}$, where the estimate for $\tau$ is the estimated ATT. 

When the treatment effect is homogeneous, the extended TWFE model in (\ref{eqn:twfe_conditional}) identifies the proper effect, with biases less than 0.1\%. However, when the treatment effect is heterogeneous, the TWFE model in (\ref{eqn:twfe_conditional}) identifies an effect with bias greater than 8\%. The bias results from misspecified effect dynamics in the model, as both $\tau$ and $\beta_{xt}$ absorb some of the effect of $\theta_i$, rather than just $\tau$. The OR and DR approaches avoid specifying a treatment effect in the model as they only model the outcome under the control exposure, making them more robust to this scenario.

\section{Testing for an Effect on Nearby PA stores not bordering Philadelphia}

\begin{table}[H]
 \centering
 \def\~{\hphantom{0}}
 \begin{minipage}{100mm}
  \caption{Effect Estimates on Non-Border Stores in zip codes within 6 miles of Philadelphia.}
\label{tab:exposure_mapping}
  \begin{tabular*}{\textwidth}{@{}r@{\extracolsep{\fill}}c@{\extracolsep{\fill}}c}
  \hline
 & SGM (million oz.) & Pharmacy (thousand oz.) \\
 \hline
 Winter & 0.08 (-0.02, 0.18) & 7.5 (2.3, 12.2) \\ 
 Spring &  0.12 (0.01, 0.26) &  2.0 (-2.6, 6.5) \\ 
 Summer &  0.20 (0.08, 0.33) &  6.2 (1.8, 10.4) \\ 
 Fall &  0.15 (0.06, 0.26) &  8.8 (4.6, 13.1) \\  
 Annual & 0.14 (0.05, 0.25) &  6.3 (3.2, 9.5) \\  
\hline
\end{tabular*}
\end{minipage}
\vspace*{18pt}
\end{table}

\section{Pre-Parallel Trends Testing}
\label{sec:appendix:pretrends}

\begin{table}[H]
 \centering
 \def\~{\hphantom{0}}
 \begin{minipage}{165mm}
  \caption{Pre-Trends Testing}
\label{tab:pretrends}
  \begin{tabular*}{\textwidth}{@{}r@{\extracolsep{\fill}}c@{\extracolsep{\fill}}c@{\extracolsep{\fill}}c@{\extracolsep{\fill}}c}
  \toprule
& \multicolumn{2}{c}{{SGM (million oz.)}} & \multicolumn{2}{c}{{Pharmacy (thousand oz.)}}  \\ [1pt]
\cline{2-3} \cline{4-5}  \\ [-6pt]
 & ATT & ATN & ATT & ATN \\
 \midrule
 Period 2  & (-0.60, -0.11) & (-0.08, 0.37) & (-9.0, 3.9) & (-6.0, 2.9) \\ 
  Period 3 & (-0.34, 0.25) & (-0.15, 0.35) & (-18.7, 8.9) & (-8.8, 11.4) \\ 
  Period 4 & (-0.62, 0.08) & (-0.16, 0.43) & (-7.6, 10.9) & (-7.0, 7.4) \\ 
  Period 5 & (-0.42, 0.25) & (-0.15, 0.37) & (-6.0, 7.0) & (-3.4, 9.6) \\ 
 Period 6  & (-0.32, 0.41) & (-0.35, 0.28) & (-8.4, 11.1) & (-2.7, 13.9) \\ 
 Period 7 & (-0.23, 0.63) & (-0.25, 0.30) & (-8.0, 13.2) & (-0.4, 14.6) \\ 
 Period 8 & (-0.53, 0.29) & (-0.28, 0.42) & (-11.3, 13.3) & (0.7, 14.2) \\ 
 Period 9 & (-0.96, 0.07) & (-0.22, 0.45) & (-10.0, 7.6) & (-1.9, 14.7) \\ 
 Period 10 & (-0.81, 0.03) & (-0.19, 0.38) & (-9.8, 9.0) & (0.5, 13.1) \\ 
 Period 11 & (-0.58, 0.07) & (-0.23, 0.25) & (-12.8, 2.9) & (-9.4, 7.0) \\ 
 Period 12 & (-0.64, 0.28) & (-0.24, 0.13) & (-18.6, -0.3) & (-6.7, 5.6) \\ 
 Period 13 & (-0.49, 0.20) & (-0.32, 0.20) & (-29.8, 0.9) & (-28.0, 6.6) \\ 
\bottomrule
\end{tabular*}
\end{minipage}
\vspace*{18pt}
\end{table}

\section{Generating Confidence Intervals}
\label{sec:appendix:ci}

In our real data analyses and simulations, we employ a bootstrap approach to generate CIs, which we find quite beneficial in our work. First, it is flexible to the data modeling approach and captures the uncertainty in our estimates due to nuisance function estimation. Second, the approach allows us to seamlessly estimate additional, complex estimands like the multiplicative estimand presented in Appendix~\ref{sec:appendix:mult_effect}. Finally, by re-sampling a store’s entire observed vector, bootstrapping automatically accounts for correlation between time-specific effect estimates when aggregating for seasonal or annual estimates.

Still, the approach may be unstable in studies with small samples and computationally demanding in others.  (Sant'anna and Zhao (2020)) provide parametric variance estimators for consistent CI estimation under strict model assumptions and as long as one of the nuisance functions is correctly specified.  However,  it is not straightforward to derive a formula for the variance in our setting with multiple time-specific effects without assuming that the time-specific effects are independent and normally distributed. Lastly, as the stratified bootstrap approach can improve stability in finite samples by limiting extreme samples, a Bayesian Bootstrap approach may similarly help while avoiding manual definition of strata of interest. 

A comparison of CI lengths from these three approaches when estimating annual effects is provided in Table~\ref{tab:confint_lengths}. The parametric variance approach is notably tighter than either of the bootstrap variance approaches due to the additional assumptions on model form and effect independence over time, which may not hold in practice. Our proposed stratified bootstrap approach produces narrower CIs than the Bayesian bootstrap, except for the Pharmacy ATT. By stratifying on region, we necessitate consistent sample sizes in each exposure group group, which may explain the tighter confidence bounds compared to the potentially more variable exposure group sample sizes generated by the Bayesian bootstrap. However, the Bayesian bootstrap does outperform the standard bootstrap, which often fails due to generated samples with zero or close to zero exposure group sizes and thus is not shown.

\begin{table}[H]
 \centering
 \def\~{\hphantom{0}}
 \begin{minipage}{165mm}
  \caption{A comparison of confidence interval lengths for annual effects}
\label{tab:confint_lengths}
  \begin{tabular*}{\textwidth}{@{}r@{\extracolsep{\fill}}c@{\extracolsep{\fill}}c@{\extracolsep{\fill}}c@{\extracolsep{\fill}}c}
  \toprule
 & \multicolumn{2}{c}{{SGM (million oz.)}} & \multicolumn{2}{c}{{Pharmacy (thousand oz.)}}  \\ [1pt]
\cline{2-3} \cline{4-5}  \\ [-6pt]
 & ATT & ATN & ATT & ATN \\
 \midrule
  Stratified Bootstrap & 1.41 & 0.91 & 21.3 & 33.3 \\ 
 Parametric Variance & 0.41 & 0.28 & 7.0 & 11.4 \\ 
 Bayesian Bootstrap & 1.62 & 1.27 & 21.2 & 41.5 \\
\bottomrule
\end{tabular*}
\end{minipage}
\vspace*{18pt}
\end{table}

\section{Estimation of a Relative Effect}
\label{sec:appendix:mult_effect}

In the real data analysis, we noted seasonal trends in the effect of the Philadelphia Beverage Tax. While these trends may be due to seasonal or temporal patterns in consumer behavior, they may also result from trends in volume beverage sales. For example, the tax may more consistently affect the \textit{percentage} of sales rather than the raw volume of sales at a given store over time. As such, focusing solely on an additive effect may paint an incomplete picture of the effect of the tax policy. In such a case, it is of interest to estimate a relative sales effect, e.g., 
\begin{equation}
ATT_{rs} := E[Y_1^{(1,0)}|g(\mathbf{A})=(1,0)] / E[Y_1^{(0,0)}|g(\mathbf{A})=(1,0)]
\label{eqn:att_rs}
\end{equation}
and the respective analogy for $ATN_{rs}$. As these are relative effects, they may also be useful when comparing effects between regions as in Section 4.3.

In order to identify such effects, we note that the numerator is observed and can be estimated with a sample mean, 
$$\tau^{dr}_{rs,1} = \mathbb{E}_n[R Y_{1}]$$
Noting that the difference of the numerator and denominator in (\ref{eqn:att_rs}) equals the additive treatment effect, the same IF estimator and identifying assumptions can be used to estimate the denominator term. Rearranging terms, we see:
$$\tau^{dr}_{rs,0} = \mathbb{E}_n[R Y_{1} - w((Y_{1}-Y_{0}) - \mu_{\Delta}(\mathbf{X}))]$$
is a doubly-robust estimate for the denominator. Our estimator for the multiplicative effect is then $\tau^{dr}_{rs} = \tau^{dr}_{rs,1} / \tau^{dr}_{rs,0}$. Since this estimator relies on a division of the two components, the bootstrapping method becomes especially helpful to estimate confidence intervals. 

The estimates for the Philadelphia and Border county regions are summarized in Table~\ref{tab:mult_estimates}. We estimate the annual $ATT_{rs}$ for SGMs as 0.46 (95\% CI: (0.37, 0.60)) and for pharmacies as 0.85 (95\% CI: (0.78, 0.92)), corresponding to 54\% and 15\% reductions in sales respectively. We estimate the annual $ATN_{rs}$ for SGMs as 1.44 (95\% CI: (1.28, 1.59)) and for pharmacies as 1.39 (95\% CI: (1.26, 1.54)), corresponding to 44\% and 39\% increases in sales respectively. The seasonal estimates display a similar pattern to those of the additive effect, which may suggest that the tax is influenced more by temporal patterns in consumer behavior than by the temporal patterns in sweetened beverage sales. 

\begin{table}[H]
 \centering
 \def\~{\hphantom{0}}
 \begin{minipage}{165mm}
  \caption{Relative Effect Estimates}
\label{tab:mult_estimates}
  \begin{tabular*}{\textwidth}{@{}r@{\extracolsep{\fill}}c@{\extracolsep{\fill}}c@{\extracolsep{\fill}}c@{\extracolsep{\fill}}c}
  \toprule
 & \multicolumn{2}{c}{{SGM (million oz.)}} & \multicolumn{2}{c}{{Pharmacy (thousand oz.)}}  \\ [1pt]
\cline{2-3} \cline{4-5}  \\ [-6pt]
 & ATT & ATN & ATT & ATN \\
 \midrule
Winter & 0.54 & 1.35 & 0.91 & 1.36 \\ 
 & (0.47, 0.64) & (1.23, 1.47) & (0.85, 1.00) & (1.22, 1.50) \\ 
 Spring & 0.42 & 1.42 & 0.82 & 1.36 \\ 
 & (0.34, 0.56) & (1.24, 1.58) & (0.74, 0.89) & (1.23, 1.49) \\ 
 Summer & 0.43 & 1.46 & 0.78 & 1.37 \\ 
 & (0.33, 0.58) & (1.29, 1.63) & (0.71, 0.86) & (1.23, 1.52) \\ 
 Fall & 0.44 & 1.51 & 0.87 & 1.47 \\ 
 & (0.34, 0.59) & (1.33, 1.68) & (0.78, 0.96) & (1.31, 1.65) \\ 
 Annual & 0.46 & 1.44 & 0.85 & 1.39 \\ 
 & (0.37, 0.60) & (1.28, 1.59) & (0.78, 0.92) & (1.26, 1.54) \\ 
\bottomrule
\end{tabular*}
\end{minipage}
\vspace*{18pt}
\end{table}

\section{Simulation parameter settings}
\label{sec:appendix:sim_param}

For all scenarios, we set $\beta^{(T)} = (0.25, 0.4, -0.5, -0.35, 0.2)$, $\beta^{(N)} = (-0.3, -0.5, -0.5, 0.5, 0.3)$, $\gamma_t = -0.5$, $\tau_m^{(ATT)} = -2$ $\forall m$, and $\tau_m^{(ATN)} = 1$ $\forall m$. Additionally, we set $\alpha_{g_i(\mathbf{A})} = 1$ and $6$ for the ATT control and exposed groups, and $0$ and $2$ for the ATN control and exposed groups.

For scenario 1, we set $\gamma_m = \{(0.5, 0.7, 0.8, 1, 1.1, 1.2, 1.4, 1.5, 1.6, 1.4, 1.1, 0.9, 0.8)\}$ and $\lambda_{0m} = \{(0.8, 0.85, 0.9, 0.95, 1, 1.05, 1.1, 1.15, 1.2, 1.15, 1, 0.95, 0.9) \}$. For scenario 2, we set $\gamma_m = \{(0.5, 1)\}$ and $\lambda_{0m} = \{(0.95, 1.05) \}$. For scenario 3, we set $\gamma_m = \{ (0.5, 1, 1.5, 1)\}$ and $\lambda_{0m} = \{ (0.85, 1, 1.1, 0.95) \}$. We set $\lambda_{1m} = 2*\lambda_{0m}$ for each scenario.

\section{Model Covariates}
\label{sec:appendix:covs}

The covariates chosen for the various nuisance function models in the real data analysis are given here. For the outcome model for the ATT SGM study in this study, we use the average house value in the zip code (house value), an indicator for mass merchandiser, and pre-tax sales of taxed beverages. For the propensity score model in this study, we use a pre-tax weighted price of taxed beveraged (weighted price), percent of the store's zip code identified as White (percent White), and the average income per household in the zip code (income). 

For the outcome model for the ATN SGM study, we use weighted price, an indicator for mass merchandiser, pre-tax sales, house value, percent white, and an interaction between the weighted price and mass merchandiser status. For the propensity score model in this study, we use weighted price, percent White, and house value. 

For the outcome model for the ATT Pharmacy study, we use percent White, income, and pre-tax sales. For the propensity score model in this study, we use a weighted price, percent White, and income. In subgroup analyses of this study, we reduced the propensity score model covariate set to percent white and income to adjust for the smaller sample sizes. 

For the outcome model for the ATN SGM study, we use weighted price, income, and pre-tax sales. For the propensity score model in this study, we use weighted price, percent White, house value, and an interaction between weighted price and house value. For the propensity score model in subgroup analyses of this study, we used weighted price, percent white, and house value for the group with low available traffic and sales from adjacent taxed zip codes and percent Black and house value for the group with high measures.

\section{Additional Simulation Results}
\label{sec:appendix:sim_results}

In Table~\ref{tab:sims_extra}, we provide simulation results for scenario (3), where $n=500$ and $n_m=4$. The results show consistent conclusions to our other scenarios, with asymptotic properties expectedly stronger than scenario (1), where $n=250$ but weaker than scenario (2), where $n=2000$.

\begin{sidewaystable}
 \centering
 \def\~{\hphantom{0}}
% \begin{minipage}{225mm}
  \hsize\textheight
  \caption{Simulation results continued.}
\label{tab:sims_extra}
\hskip-12pt\begin{tabular*}{\textheight}{@{}l@{\extracolsep{\fill}}c@{\extracolsep{\fill}}c@{\extracolsep{\fill}}c@{\extracolsep{\fill}}c@{\extracolsep{\fill}}c@{\extracolsep{\fill}}c@{\extracolsep{\fill}}c@{\extracolsep{\fill}}c@{\extracolsep{\fill}}c@{\extracolsep{\fill}}c@{\extracolsep{\fill}}c@{\extracolsep{\fill}}c@{\hskip12pt}}
  \toprule
 & \multicolumn{4}{c}{{Bias ($\%$)}} & \multicolumn{4}{c}{{Std. Err}} & \multicolumn{4}{c}{{Coverage ($\%$)}} \\ [1pt]
\cline{2-5} \cline{6-9} \cline{10-13} \\ [-6pt]
& & & & & & & & & & & & \\ [-3pt]
Scenario & TWFE & IPW & OR & DR & TWFE & IPW & OR & DR & TWFE & IPW & OR & DR
\\ 
 \midrule
 \multicolumn{1}{l}{\textit{ATT}} & & & & & & & & & \\
 \midrule
 3a & -11.976 & 0.128 & 0.713 & 0.139 & 0.423 & 0.057 & 0.353 & 0.059 & 90.5 & 93.1 & 97.1 & 93.8 \\
 3b & -43.696 & 0.152 & -1.364 & 0.187 & 0.436 & 0.058 & 0.33 & 0.059 & 45.4 & 92.5 & 94.8 & 94.0 \\
 3c & -5.75 & -9.288 & -0.239 & -2.410 & 0.394 & 0.279 & 0.620 & 0.451 & 93.7 & 88.7 & 95.0 & 92.8 \\
 3d & 18.497 & 27.495 & 34.590 & 32.256 & 0.400 & 0.253 & 0.352 & 0.288 & 83.2 & 42.4 & 42.9 & 34.1 \\ 
 \midrule
 \multicolumn{1}{l}{\textit{ATN}} & & & & & & & & & \\
 \midrule
 3a & -51.2 & -0.080 & 0.005 & -0.166 & 0.357 & 0.048 & 0.229 & 0.051 & 72.2 & 98.7 & 96.7 & 98.4 \\
 3b & -23.631 & -0.040 & -24.953 & -0.021 & 0.369 & 0.044 & 0.590 & 0.048 & 89.4 & 99.0 & 94.5 & 98.9 \\
 3c & 50.125 & -30.159 & -0.983 & -7.144 & 0.353 & 0.247 & 0.249 & 0.304 & 69.9 & 77.8 & 95.9 & 93.1 \\
 3d & -1.775 & -51.697 & -59.399 & -46.790 & 0.352 & 0.243 & 0.426 & 0.330 & 94.6 & 41.8 & 49.0 & 55.6 \\ 
\bottomrule
\end{tabular*}
%\end{minipage}
\end{sidewaystable}

\end{document}